\documentclass{emulateapj}
\usepackage{epsfig}

\begin{document}

\title{The Detection of Inside-out Disk Growth in M33}

\author{Benjamin F. Williams\altaffilmark{1},
Julianne J. Dalcanton\altaffilmark{1},
Andrew E. Dolphin\altaffilmark{2},
Jon Holtzman\altaffilmark{3},
Ata Sarajedini\altaffilmark{4}
}
\altaffiltext{1}{Department of Astronomy, Box 351580, University of Washington, Seattle, WA 98195; ben@astro.washington.edu; jd@astro.washington.edu}
\altaffiltext{2}{Raytheon, 1151 E. Hermans Road, Tucson, AZ 85706; dolphin@raytheon.com}
\altaffiltext{3}{Department of Astronomy, New Mexico State University, Box
30001, 1320 Frenger St., Las Cruces, NM 88003; holtz@nmsu.edu}
\altaffiltext{4}{Department of Astronomy, University of Florida, 211 Bryant Space Science Center, PO Box 112055, Gainesville, FL, 32611; ata@astro.ufl.edu}


\begin{abstract}

We present resolved stellar photometry of 4 fields along the major
axis of the M33 disk from images taken with the Advanced Camera for
Surveys aboard the Hubble Space Telescope.  The photometry provides a
detailed census of the red clump in all fields and reaches the ancient
main sequence in the outermost field.  Through detailed modeling of
the color-magnitude diagrams, we find that the percentage of the
stellar mass formed prior to $z=1$ changes from 71$\pm$9\% in the
inner-most field to 16$\pm$6\% in the outermost field.  The disk shows
a clear trend of increasing scale-length with time, evolving from $r_s
= 1.0\pm0.1$ kpc 10 Gyr ago to $r_s = 1.8\pm0.1$ kpc at times more recent
than 5 Gyr ago, in agreement with analytical predictions for disk
growth.  Beyond the disk truncation radius, however, the stellar
density profile steepens with time and the age gradient reverses, in
agreement with recent simulations.  The late and slow growth of the
stellar disk may be due in part to the low mass of M33.

\end{abstract}

\section{Introduction}

The scale-lengths of galaxy disks are thought to increase with time
due to late time accretion of gas at large radii and exhaustion of gas
in the centers of galaxies \citep[][and many
others]{larson1976,matteucci1989,burkert1992,chiappini1997,naab2006}.
This process, which has been called ``inside-out'' growth, reproduces
many of the physical \citep[e.g.,][]{bouwens1997} and chemical
properties of disks.  Observations of color gradients in spiral
galaxies suggest that the mean ages of the stars decrease from the
inside-out \citep[e.g.,][]{macarthur2004}, and observations of
extended ultra-violet disks show that very recent star formation is
prevalent in the outer regions of disks \citep[e.g.,][]{boissier2008}.
While these trends are consistent with inside-out growth, broadband
colors can be biased by variations in star formation, dust extinction,
and metallicity, while ultra-violet emission probes only the current
disk properties.  A more direct approach to detecting the products of
long-term inside-out disk growth is through measurements of the age
distribution of the resolved stellar populations.

The inside-out growth of stellar disks may be most easily detected in
low-mass disks. If massive disks formed earlier than low-mass disks
\citep[i.e. ``down-sizing''][and many
others]{cowie1996,lilly2003,neistein2006,fontanot2009}, the stars in
massive disks have had more time to be mixed by radial redistribution
processes \citep[e.g.,][]{roskar2008}.  Therefore detecting the
signature of inside-out growth using stellar populations may be
simplified in low-mass disks. 

M33 is the nearest example of a low-mass disk galaxy.  It is also a
bulge-less system \citep{mclean1996}, so that there is no confusion
between disk and bulge populations.  If the stellar disk of M33 grew
from the inside-out, the age of the stellar populations should vary
with radius, with a higher percentage of ancient stars in the inner
disk. Indeed, ground-based observations of M33's bright stellar
populations suggest that its outer disk may have formed at late epochs
and may even be ongoing \citep{davidge2003,block2004,rowe2005}.  On
the other hand, M33 has a truncated disk \citep[at r$\sim$8
kpc,][]{ferguson2007} and the age of populations of the outer
disk/halo show an age increase with radius based on HST/ACS data
\citep{barker2007,barker2007b}, in contrast to the expectations of
inside-out growth.

In this letter, we use data from the Hubble Space Telescope (HST)
Advanced Camera for Surveys (ACS) to study the age distribution of the
stellar populations of the M33 disk as a function of galactocentric
radius.  These distributions show a clear trend: regions nearer to the
center formed a higher fraction of their stars at earlier times
leading to a steady growth in the disk scale length.  Our measurements
therefore provide strong observational support for inside-out disk
growth.  

More rigorous checks of this analysis and more detailed studies of
these fields will be provided in a future full-length paper (Holtzman
et al. 2009, in preparation). We assume a distance of 800 kpc
\citep{lee2002} for conversions of angular measurements to physical
distances.  We assume an inclination of 56$^\circ$
\citep{corbelli1997} for de-projections. We adopt a
WMAP\footnote{http://lambda.gsfc.nasa.gov/product/map/dr2/params/lcdm\_wmap.cfm}
cosmology for all conversions between time and redshift.

\section{Data Acquisition and Processing}

\begin{figure*}[!t]
\centerline{\psfig{file=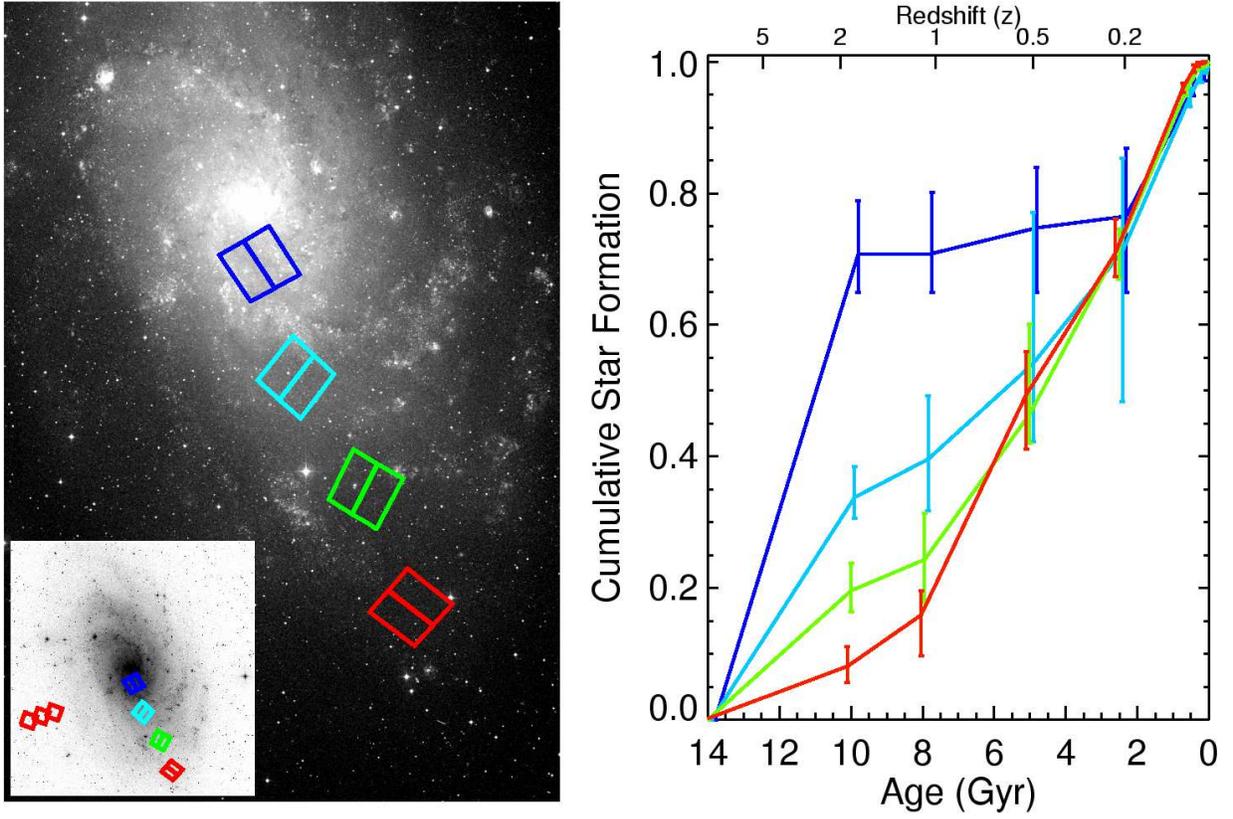,width=6.5in,angle=0}}
\caption{{\it Left:} A digitized sky survey image of M33.  Overplotted
  in color are our 4 HST/ACS fields.  The inset image shows the
  locations of the 3 \citet{barker2007} fields as red squares.  {\it
  Right:} The cumulative fraction of stars formed in the four inner
  fields as a function of time.  The colors of the lines are matched
  to the colors of the field outlines in {\it left}.}
\label{all}
\vspace{0.3cm}
\end{figure*}

We obtained HST/ACS imaging of 4 inner fields along the major axis of
the M33 disk as part of the HST GO program 10190 (see left panel of
Figure~\ref{all}).  The observations were performed from 09-Sep-2004
to 20-Feb-2005 through the F606W (wide-$V$) and F814W ($I$ equivalent)
filters.  The total exposure time in the innermost field was 6260 s
and 6482 s in F606W and F814W, respectively.  The total exposure time
for each of the other 3 fields was 21140 s and 26300 s in F606W and
F814W, respectively.

We also analyzed HST/ACS imaging of 3 outer fields along the minor
axis of the M33 disk (GO-9479), which have been previously studied by
\citet{barker2007}.  Our reanalysis of these fields allowed us to
extend our radial profiles to larger radii and provided a consistency
check between our analysis technique and those used for previous work.

\begin{figure*}[!t]
\centerline{\psfig{file=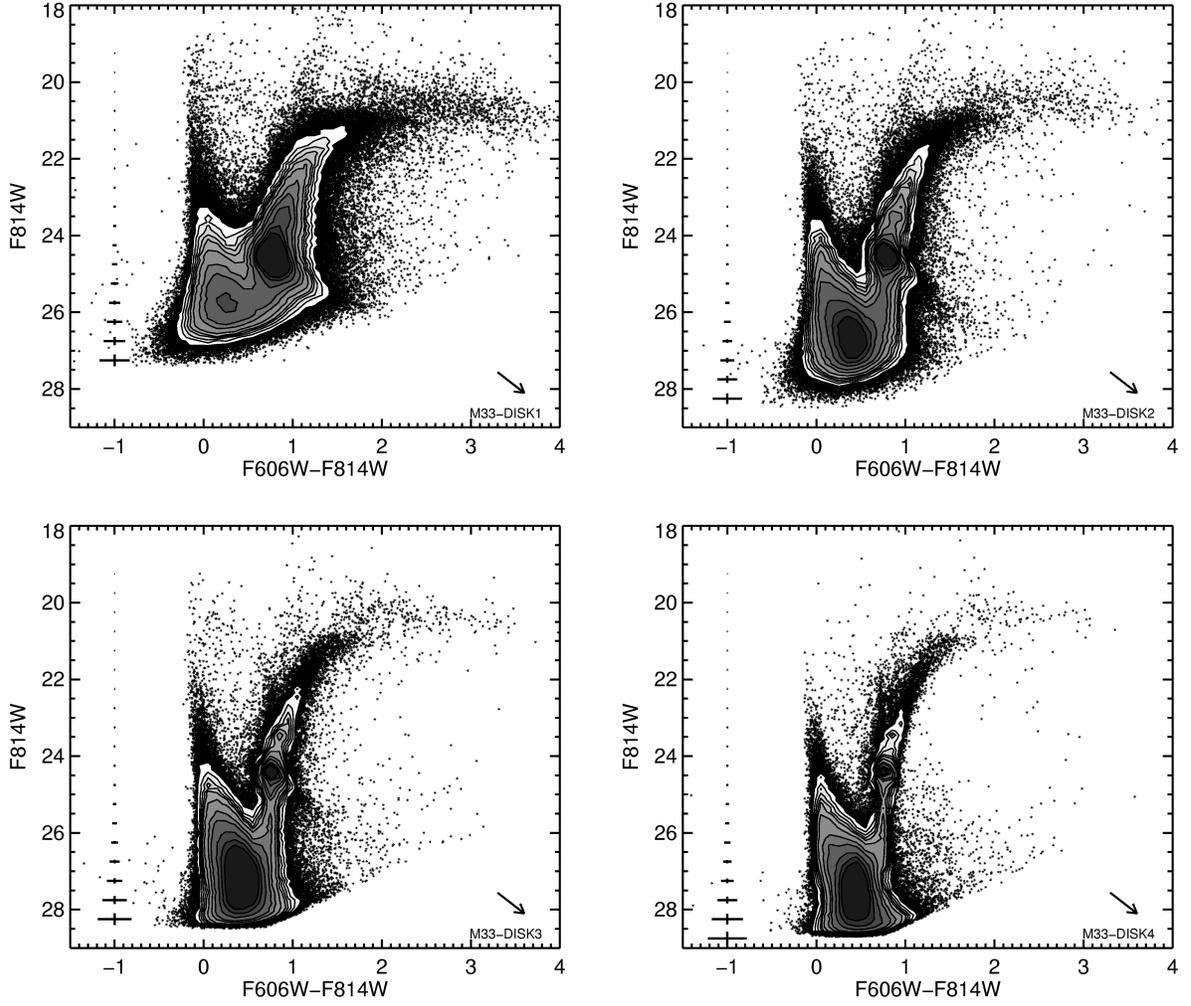,width=6.5in,angle=0}}
\caption{{\it Upper Left:} The F606W-F814W CMD of the innermost field.
  Contour levels indicate density of points where the plot would be
  saturated (levels are 1,1.5,2,2.5,3,4,6,8,12,16,20 $\times10^4$
  mag$^{-2}$).  Photometric errors as a function of F814W magnitude
  are shown on the left side.  The field name and reddening vector are
  shown in the lower-right corner.  {\it Upper Right:} Same as {\it
  upper left}, for the next field out from the center.  {\it Lower
  Left:} Same as {\it upper right}, for the next field out.  {\it
  Lower Left:} Same as {\it lower left}, but for the outermost field.
  For CMDs of the archival fields, see \citet{barker2007}.}
\label{cmds}
\end{figure*}

Resolved stellar photometry was measured using the photometric
pipeline of the ACS Nearby Galaxies Survey Treasury
\citep{dalcanton2009}.  The pipeline uses the DOLPHOT
\citep{dolphin2000} software package to measure point spread function
photometry and to determine errors and completeness as a function of
color and magnitude using artificial star tests
\citep{dalcanton2009,williams2008}.  The final stellar catalogs
contained 414780, 386894, 407741, and 309249 stars, from the innermost
field to the outermost field, respectively.  The resulting
color-magnitude diagrams (CMDs) are shown in Figure~\ref{cmds}.  Our
measurements of the \citet{barker2007} fields yielded 13522, 3958, and
1548 stars from the innermost field to the outermost field.  The
resulting CMDs were equivalent to those shown in \citet{barker2007}.

The CMDs shown in Figure~\ref{cmds} qualitatively show the relative
number of main sequence stars to giants increasing with the distance
from the center of M33, suggesting decreasing stellar ages with
increasing radius.  To quantify this trend, we measured the star
formation history (SFH) of each of the 4 fields using the software
package MATCH \citep{dolphin2002}, which finds the best-fitting
distribution of stellar ages and metallicities that reproduces the
observed CMD assuming a \citet{salpeter1955} initial mass function
(IMF) and the stellar evolution isochrones of \citet{girardi2002}
including recent updates \citep{marigo2008}.  The details of the
fitting method used were the same as those employed in
\citet{williams2008}, where results at old ages were binned to low
time resolution to reduce errors and avoid over-interpreting details
that could be due to deficiencies in the stellar evolution
models. This technique for recovering the SFH from the CMD is
well-tested and is found to be robust against differences in stellar
evolution models and fitting techniques when the distance and mean
extinction to the field are allowed to be fitted as free parameters.

Uncertainties were determined by fitting the CMDs with a range of
assumed distance and mean reddening values.  These uncertainties were
then added in quadrature to the standard deviation calculated from
Monte Carlo tests where the observed CMD was resampled and refitted
one hundred times.

\begin{figure*}
\centerline{\psfig{file=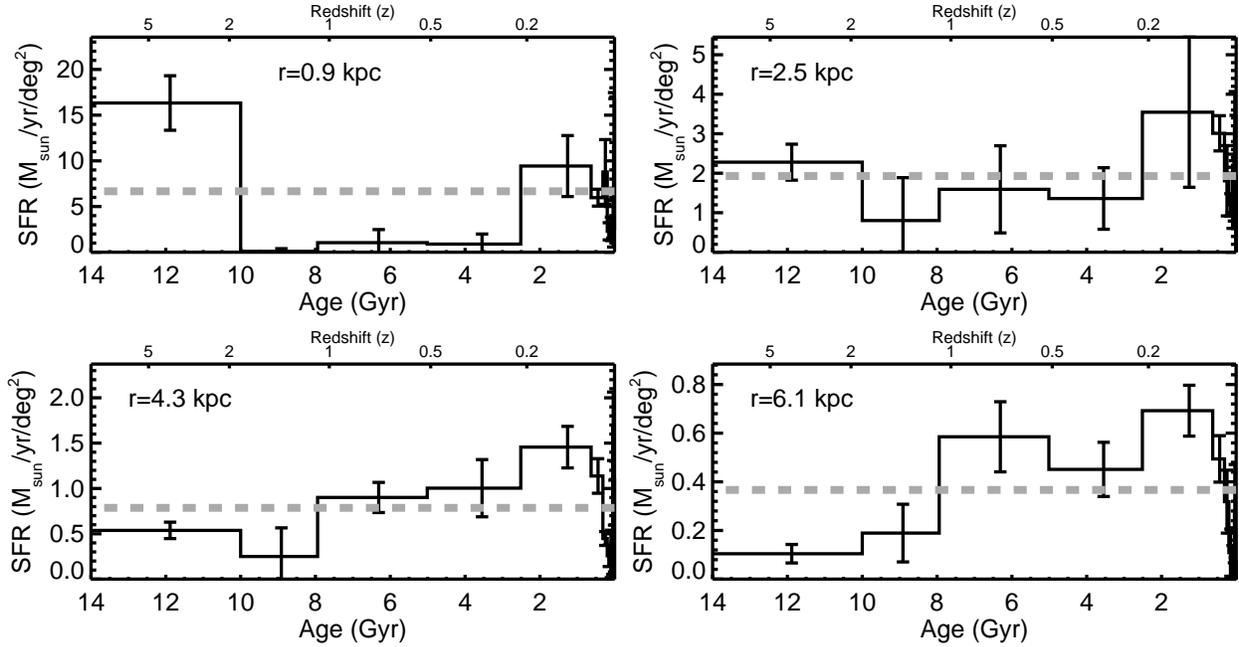,width=6.5in,angle=0}}
\caption{{\it Upper Left:} The star formation rate as a function of
time of the innermost field.  A thick, gray, dashed line denotes the
mean rate for the field. {\it Upper Right:} Same as {\it upper left},
but for the next field out from the center.  {\it Lower Left:} Same as
{\it upper right}, but for the next field out.  {\it Lower Left:} Same
as {\it lower left}, but for the outermost field. For SFHs of the
archival outer fields, please see \cite{barker2007}.}
\label{sfhs}
\end{figure*}

A constant mean extinction was applied to all model stars, using the
ACS extinction coefficients from \citet{girardi2008}. Greater
reddening was assumed for stars with ages $<$100 Myr, but these ages
are not relevant to this study.  As a first-order attempt to account
for differential reddening, we applied a uniform spread in $A_V$ to
the model photometry of 0.6, 0.4, 0.2, and 0.0 magnitudes for the
innermost to outermost fields (respectively).  These values improved
the quality of the model fits compared to fits with a single uniform
reddening value for each field. However, differential reddening did
not significantly alter the inferred SFHs (see \S~3.4), which are
shown in Figure~\ref{sfhs}.

\section{Results} 

\subsection{Star Formation Histories}

The M33 disk is naturally divided into two components: the inner disk
and outer disk.  These components are divided by a break in the
exponential density profile located at a radius of $\sim$8 kpc
\citep{ferguson2007}.  The radius of this truncation point is located
between our outermost field and the innermost \citet{barker2007}
field.

The SFHs of the fields (see Figure~\ref{sfhs}) show that the bulk of
the stars in each ACS field formed at different times, as can be seen
most easily by looking at the cumulative fraction of stars formed
during each time bin.  This value is plotted as a function of age for
all 4 inner fields in the right panel of Figure~\ref{all}.  The plot
clearly shows that, while the majority of the stars near the center of
the disk had formed by $z=1$, the bulk of the stars farther out in the
disk formed later.

The growth is also shown in Figure~\ref{sl}, which plots the
reconstructed stellar mass density profile of the M33 disk at 5
different epochs.  This plot shows that most of the stellar disk
growth occurred at radii of $\sim$3--7 kpc, rather than at the center,
leading to an increase in the scale length of the inner disk (r$<$8
kpc) with little change in the central stellar surface density.  In
contrast, while the density profile of the inner disk has flattened as
the disk has aged, that of the outer disk has steepened.

\subsection{Implications for the Inner Disk}

\begin{figure*}
\centerline{\psfig{file=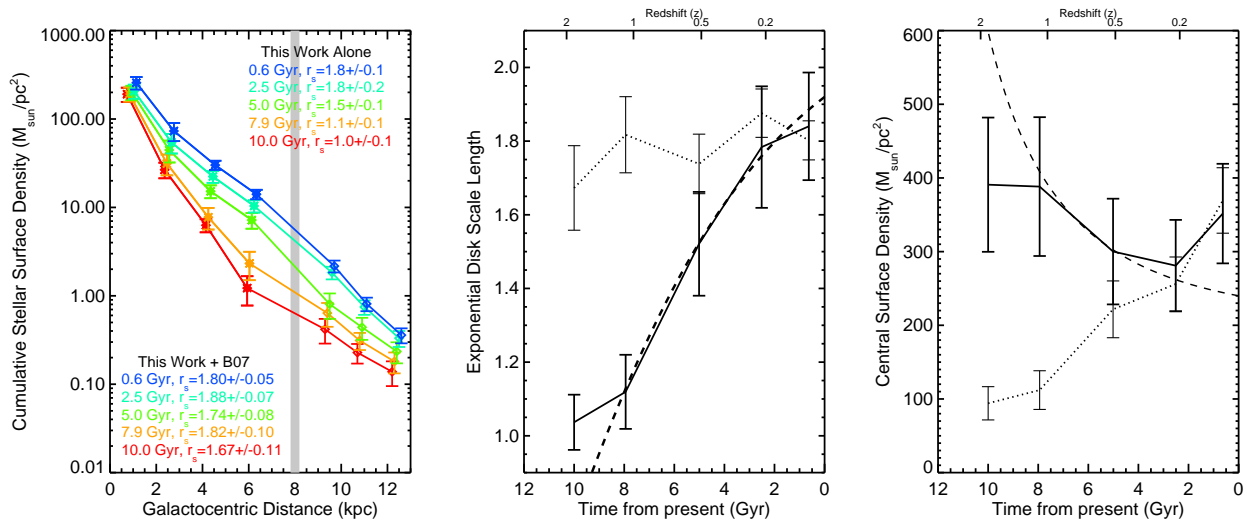,width=6.5in,angle=0}}
\caption{{\it Left:} M33 stellar surface densities as a function of
  de-projected radius as computed from our measured SFHs (stars) and
  from our re-analysis of the \citet{barker2007} fields (diamonds).
  Different colors show the total stellar densities that would be
  observed at different lookback times, noted in the corners. The
  absolute mass density values on the ordinate can shift based on the
  IMF.  The best-fit exponential scale-length for each epoch using
  just the fields inside the disk break is listed in the upper-right.
  Those using all of the fields are listed in the lower-left.  Data
  points have been offset from one another by 0.1 kpc to avoid
  overlapping error bars. The vertical gray line marks the disk break
  measured by \citet{ferguson2007}. {\it Center:} The disk scale
  length as a function of age inside the disk break (solid line) and
  from all fields (dotted line). The dashed line shows the scaling
  relation of \citet{mo1998} normalized to our 5~Gyr measurement. {\it
  Right:} The central surface density as a function of age as measured
  from fields inside the disk break (solid line) and from all fields
  (dotted line). The dashed line shows the scaling relation of
  \citet{mo1998} normalized to our 5~Gyr measurement.}
\label{sl}
\end{figure*}

We have fit the inner disk (r$<$7 kpc) with an exponential profile and
plotted the results in the central panel of Figure~\ref{sl}.  The
scale length of M33's disk has increased by a factor of $\sim$2 from
10 Gyr ago to the recent past.  Most of this stellar disk growth
occurred from 10 Gyr to 5 Gyr ago.  This growth is similar to the
predictions of \citet{mo1998}, which are based on simple scaling
models for the evolution of the virial mass and virial radius.

The current stellar mass scale-length is longer than that of the
$K$-band light \citep[][5.8$'$ in $K$ vs. 7.7$'$ here]{regan1994},
suggesting a changing M/L$_K$ with radius. This discrepancy has been
seen in a dynamical estimate of the scale length of M33's mass
\citep{ciardullo2004}. However, the $K$-band scale length is similar
to the scale length we calculate for stellar populations older than 5
Gyr.

The right panel of Figure~\ref{sl} shows the extrapolated central
surface density of the exponential fits to the stellar mass density
profile.  The uncertainties are significant, and the data are
consistent with no evolution in the central surface density of the
stellar disk (solid line).  We compare the central surface densities
with the expectations from the scaling laws of \citet{mo1998};
however, those laws track only the mean distribution of the baryons
and formally predict a decrease in surface density with time.  This
prediction is not easily compared with our measurement for the {\it
stellar} disk, for which such a decrease is likely to be unphysical.

Based on Figure~\ref{sl}, we conclude that most of the stars outside
of the innermost field formed since $z=1$.  However, this conclusion
differs from what has been seen in the outer disk of more massive
spirals, such as M31 \citep{brown2006}, M81 \citep{williams2008}, and
the Milky Way thick disk, in which most of the stars appear to have
formed prior to $z=1$.  This early formation of the disks of massive
spirals is also seen in redshift surveys, which are sensitive to
equally massive spirals beyond $z=1$
\citep{lilly1998,ravindranath2004,papovich2005,sargent2007,melbourne2007}.
Such surveys are not sensitive to low-mass disks like M33.  The
difference between our results for M33 and more massive disks may be
due to the fact that such low-mass galaxies are thought to perform
most of their star formation at later epochs, an effect that has been
deemed ``down-sizing.''

\subsection{Comparison to the Outer Disk}

There is a striking difference between the age gradient seen in our
data for the inner disk and that seen in the far outer disk by
\citet{barker2007}.  Within their series of outer disk fields, their
measurements show the mean age of the stars increasing with radius.
In contrast, our analysis shows the opposite effect within the break
radius, marked by the vertical gray line in the left panel of
Figure~\ref{sl}. Similar differences between the inner and outer disk
are seen in current simulations of disk formation, where radial mixing
redistributes older stars to larger radii.  This mixing produces a
break in the exponential disk profile, inside of which stellar ages
decrease with increasing radius and outside of which stellar ages
increase with increasing radius \citep{roskar2008}.  Although this
radial mixing scatters stars into the outer disk, the majority of the
stars in the inner disk remain close to their formation radius.  Thus,
while mixing produces the break and age gradient in the outer regions,
it is not sufficient to erase the radial age gradient established
during the formation of the inner disk.

Our independent reduction of the \citet{barker2007} data and
derivation of the SFH reproduces their observed increase in mean age
with radius in the outer disk.  The resulting surface densities of
these outer fields are included in Figure~\ref{sl}.  Fits of a single
exponential to the full broken exponential (see center panel of
Figure~\ref{sl}) average the inner and outer disk gradient and thus
show no significant change in scale-length with age.  This exercise
demonstrates the importance of measuring profiles of the disk inside
and outside of the disk break separately.

Finally, our results show some very old stars at all radii in the
disk.  However, in disk formation and evolution simulations, radial
redistribution places stars of the oldest ages throughout the disk,
including much larger radii than where they were formed
\citep{roskar2008}, suggesting that the oldest stars seen at larger
radii could have been formed closer to the galaxy center. The oldest
stars may also have been accreted. In any case, the inner disk
apparently had a shorter scale-length in the past, and the trends seen
in mean age and scale-length in our measurements mimic those seen in
simulations and semi-analytic models of inside-out growth.

\subsection{Reliability Tests}

Because our conclusions rely strongly on the reliability of our
technique to measure SFHs from resolved stellar photometry, we have
performed several tests to check the robustness of our measurements.
Our conclusions draw upon the age distribution of stars with ages of
2-12 Gyr, resolved into 4 time bins.  We performed multiple binnings
to check the sensitivity of our results to binning scheme and verified
that the same age gradient was present even for larger more
conservative time bins than those finally adopted.

While such high time resolution in the 2-12 Gyr range is more robust
at the depths of the outer 3 major axis fields, the ability to
distinguish between stars with ages of 5 and 10 Gyr may be less
reliable in the shallower innermost field.  We therefore calculated
the scale length of the inner disk as a function of time excluding the
innermost field and found the trend to be equivalent to that measured
when the innermost field is included.

We also tested the effects of forcing the distance to be the same in
each fit by assigning a fixed distance of $(m-M)_0$=24.6 to all of the
fields. This test decreased the scale lengths at all epochs, but left
the trend the same.

The CMDs contain stars with a spread in reddening, such that the red
clump appears slightly smeared along the reddening line.  To test the
effects of our assumption that differential reddening effects are less
severe in fields farther out in the disk (see \S~2), we refit our data
first assuming the same spread in reddening ($A_V=0.6$) in the
innermost 3 fields and again assuming no differential reddening in any
of the fields.  While these changes affected the resulting overall
scale lengths, causing the trend to be from 1.0$\pm$0.1 kpc to
1.7$\pm$0.2 kpc for a spread of $A_V=0.6$ and from 1.1$\pm$0.1 kpc to
2.0$\pm$0.2 kpc for no spread at all, the changes did not
significantly alter the resulting SFHs or change the trend of
increasing scale length with time.  More sophisticated tests of these
effects will be reported in the more comprehensive paper on this
dataset (Holtzman et al. 2009, in preparation).

We also checked for consistency with previous work.  Our measured SFHs
for the \citet{barker2007} fields were consistent with theirs within
the errors except for a factor of $\sim$2 offset in the normalization,
due to the IMF.  Since \citet{barker2007} assumed an IMF that flattens
at low masses, their models required lower star formation rates to
reproduce the observed CMD. When normalized to produce the same number
of stars from 1 to 100 M$_{\odot}$, the ratio of the total number of
stars produced by their adopted IMF to that produced by our adopted
IMF is $\sim$0.5.  Correcting for this offset puts our resulting SFHs
in good alignment with theirs. Changes in the IMF and mass cutoff
serve only to shift all of the surface density data in Figure~\ref{sl}
up or down.  The scale-lengths and fractional changes in central
surface density are thus unaffected by these IMF issues, unless the
true IMF varies with radius \citep{meurer2009}.

\section{Conclusions}

We have modeled the CMDs of 7 HST/ACS fields in the M33 disk at a
range of galactocentric radii.  Four of these fields are
newly-observed and lie along the major axis of the inner disk.  The
resulting stellar age distributions demonstrate that, within the inner
disk, the age of the majority of the stars decreases with increasing
distance from the galactic center.  In addition, most of the disk
stars outside of $\sim$3 kpc formed since $z=1$, later than what is
seen in more massive spiral galaxies both in stellar populations
studies and in redshift surveys.  Comparisons of the stellar
population properties in these fields with those of the outer M33 disk
as measured by \citet{barker2007} show an inversion of the radial
dependence of stellar age, providing support for some current
simulation results.  Together, these results provide strong
observational support to inside-out growth of low-mass spiral stellar
disks, down-sizing, and the significance of disk breaks in formation
models of galaxy disks.

Support for this work was provided by NASA through grants GO-10190 and
GO-10915 from the Space Telescope Science Institute, which is operated
by the Association of Universities for Research in Astronomy,
Incorporated, under NASA contract NAS5-26555.



\end{document}